\documentclass[12pt,final,journal,onecolumn]{IEEEtran}
\usepackage[english]{babel}
\usepackage{mathrsfs}
\usepackage{amssymb}
\usepackage{hyperref}

\newcommand{\Aut}{\ensuremath{\mathrm{Aut}}}
\newcommand{\Sym}{\ensuremath{\mathrm{Sym}}}
\newcommand{\Ker}{\ensuremath{\mathrm{Ker}}}

\newcommand{\supp}{\ensuremath{\mathrm{supp}}}
\newcommand{\wt}{\ensuremath{\mathrm{wt}}}
\newcommand{\Z}{\ensuremath{\mathbb{F}}}

\begin{document}
\bibliographystyle{IEEEtran}

\title{The Perfect Binary One-Error-Correcting Codes of Length $15$:
Part I---Classification}
\author{Patric R. J. {\"O}sterg{\aa}rd, Olli~Pottonen
\thanks{This work was supported in part by the Graduate School in
Electronics, Telecommunication and Automation and by
the Academy of Finland, Grant Numbers 107493 and 110196.}%
\thanks{P. R. J. {\"O}sterg{\aa}rd and O. Pottonen are with the
Department of Communications and Networking, Helsinki University of
Technology TKK, P.O.Box 3000, FI-02015 TKK, Finland (e-mail:
patric.ostergard@tkk.fi, olli.pottonen@tkk.fi)}}

\maketitle
\begin{abstract}
A complete classification of the perfect binary one-error-correcting
codes of length $15$ as well as their extensions of length $16$ is presented.
There are $5\,983$ such inequivalent perfect codes and
$2\,165$ extended perfect codes.
Efficient generation of these codes relies on
the recent classification of Steiner quadruple systems of order $16$.
Utilizing a result of Blackmore, the optimal binary one-error-correcting
codes of length $14$ and the $(15, 1\,024, 4)$ codes are also classified;
there are $38\,408$ and $5\,983$ such codes, respectively.
\end{abstract}
\begin{keywords}
classification, Hamming code, perfect binary code, Steiner system
\end{keywords}

\section{Introduction}

Consider the space $\Z^n_2$ of dimension $n$ over
the Galois field $\mathbb{F}_2 = \{0,1\}$. A binary \emph{code} of length $n$ is
a subset of $\Z^n_2$.
The \emph{(Hamming) distance} $d({\bf x}, {\bf y})$ between two codewords
${\bf x}$, ${\bf y}$ is the number of coordinates in which they differ,
and the \emph{(Hamming) weight} $\wt({\bf x})$ is the number of nonzero
coordinates. The \emph{support} of a codeword is the set of nonzero
coordinates, $\supp({\bf x}) = \{i : x_i \neq 0\}$. Accordingly,
$d({\bf x}, {\bf y}) = \wt({\bf x}-{\bf y}) = |\supp({\bf x}-{\bf y})|$.

A code has \emph{minimum distance} $d$ if $d$ is the largest integer 
such that the distance between any distinct codewords
is at least $d$. Then the balls of radius $\lfloor (d-1)/2 \rfloor$
centered around the codewords are nonintersecting, and the code is said to be
a  $\lfloor (d-1)/2 \rfloor$-error-correcting code. If these balls tile the
whole space, then the code is called \emph{perfect}. The parameters of perfect
codes over an alphabet of prime order are well known~\cite{MacWilliams77},
and perfect binary
codes exist with $d=1$; $d=n$; $d=(n-1)/2$ for odd $n$;
$d=3, n=2^m-1$ for $m\ge 2$; and $d=7, n=23$. The first three types of codes
are called trivial, the fourth has the parameters of Hamming codes,
and the last one is the binary Golay code. A perfect code with minimum distance
$d$ is also called a $\lfloor (d-1)/2 \rfloor$-perfect code.

A binary code with length $n$, minimum distance $d$, and $M$ codewords is called
a $(n, M, d)$ code. In this notation a binary $1$-perfect code is a
$(2^m-1, 2^{2^m-m-1}, 3)$ code. Two related families are the extended
and shortened $1$-perfect codes, which have parameters
$(2^m, 2^{2^m-m-1}, 4)$ and $(2^m-2, 2^{2^m-m-2}, 3)$, respectively.

Existence of binary $1$-perfect codes follows from the existence of
Hamming codes, which are the unique \emph{linear} $1$-perfect codes.
Still constructing all $1$-perfect codes is a longstanding open
problem. It makes sense to approach this issue by considering the
number of \emph{inequivalent} codes (or more formally the number of
equivalence classes). Two codes are said to be
equivalent if one is obtained from the other by permuting coordinates
and adding a constant vector; a formal definition appears in
Section~\ref{sec:prel}. 

There is trivially a unique $1$-perfect code of length $3$.
Zaremba~\cite{Zaremba52} showed that also the $1$-perfect code of
length $7$ is unique. However, already the next case of length $15$ has
until this work withstood all attempts of complete classification, although
several constructions of such codes have been published;
see the surveys~\cite{Etzion94, Heden08}.
It turns out that these results were not far from a complete classification
as for the number of codes found.
The growth of the number of $1$-perfect binary codes is double exponential in the length 
of the code, see \cite{Krotov08} for a lower bound on this number.
For an in-depth treatment of the topic of classifying
combinatorial objects, see~\cite{KO}.

The aim of the current work is to obtain a complete classification
of inequivalent $1$-perfect binary codes of length $15$.
By computer search it is here shown that their number is
$5\,983$. Also the codes obtained by extending, shortening or extending and
shortening are classified;
the numbers of $(16, 2\,048, 4)$, $(14, 1\,024, 3)$ and $(15, 1\,024, 4)$
codes turn out to be $2\,165$, $38\,408$ and $5\,983$ respectively.

In the rest of the paper we document the classification of the
extended 1-perfect
codes of length $16$, which yields classifications of the 1-perfect codes
of length $15$ and the shortened 1-perfect codes of length $14$.
In Section~\ref{sec:prel} we define some concepts and consider construction
of extended 1-perfect codes via Steiner systems.
In Section~\ref{sec:isom} we present algorithms for detecting and
rejecting equivalent codes, and in
Section~\ref{sec:results}, we take a brief look at
the results; a separate, more detailed study of the 
classified codes will appear in a separate paper~\cite{OPP09}.
Finally, in Section~\ref{sec:check} we give a consistency check for
gaining confidence in the computational results.

\section{Preliminaries and Construction}\label{sec:prel}

A permutation $\pi$ of the set $\{1,2,\ldots,n\}$ acts on codewords
by permuting the coordinates: $\pi((c_1, c_2, \ldots, c_{n})) =
(c_{\pi^{-1}(1)}, c_{\pi^{-1}(2)}, \ldots, c_{\pi^{-1}(n)})$.
Pairs $(\pi, {\bf x})$ form the
\emph{wreath product} $S_2\wr S_n$, which acts on codes as
$(\pi, {\bf x})(C) = \pi(C + {\bf x}) = \pi(C) + \pi({\bf x})$.
Two codes, $C_1$ and $C_2$, are \emph{isomorphic}
if $C_1 = \pi(C_2)$ for some $\pi$ and \emph{equivalent}
if $C_1 = \pi(C_2 + {\bf x})$ for some $\pi, {\bf x}$.

The \emph{automorphism group} of a code $C$, $\Aut(C)$, is the group
of all pairs $(\pi, {\bf x})$ such that $C = \pi(C + {\bf x})$.
Two important subgroups of $\Aut(C)$ are the \emph{group of symmetries},
\[
\Sym(C) = \{\pi : \pi(C) = C\}
\]and the \emph{kernel}
\[
\Ker(C) = \{{\bf x} : C + {\bf x} = C\}.
\]
If the code contains the all-zero word, ${\bf 0}$, then the elements of
the kernel are codewords.

A \emph{Steiner system} $S(t,k,v)$ can be viewed as a code
$S \subset \mathbb{F}^v_2$ with the property that each codeword of $S$ has weight
$k$, and for any ${\bf y} \in \mathbb{F}_2^v$ with $\wt({\bf y}) = t$,
there is a unique ${\bf x} \in S$ such that
$\supp({\bf y}) \subseteq \supp({\bf x})$. Usually Steiner systems
are defined as set systems rather than codes, but our definition is
more directly applicable for this work.
The parameter $v$ is the \emph{order} of the system.
Steiner systems $S(2,3,v)$ and $S(3,4,v)$, which are called
\emph{Steiner triple systems} and \emph{Steiner quadruple systems},
respectively, are related to 1-perfect codes in the following way.
If $C$ is a 1-perfect
binary code of length $v$ and ${\bf x} \in C$, then the codewords
of $C + {\bf x}$ with weight $3$ form a Steiner triple system of order $v$.
Similarly, if $C$ is an extended 1-perfect binary code and ${\bf x} \in C$, then
the codewords of $C + {\bf x}$ with weight $4$ form a Steiner quadruple system.
These systems are, respectively, the \emph{neighborhood triple system} and
\emph{neighborhood quadruple system} associated with the code and the
codeword.

As a starting point for the classification of the extended
1-perfect binary codes of length 16, we have the classification~\cite{sqs16}
of Steiner quadruple systems of order 16; there are $1\,054\,163$ such
designs.  We want to find, for each $S(3,4,16)$, all extended
1-perfect binary codes in which it occurs. This can be done by
puncturing any coordinate, augmenting the resulting code to 1-perfect
codes in all possible ways, and finally extending every resulting code
with a parity bit.

When augmenting a set of codewords to a 1-perfect code, we consider a
1-perfect code as a set of balls with radius one that 
form a partition of the ambient space. Accordingly, finding a code
(with specified codewords) is a special case of
the \emph{exact cover problem},
where we are given a set $S$ and a collection $U$ of its subsets, and the
task is to form a partition of $S$ by using sets in $U$.
Let the set $Q$ contain the codewords obtained by puncturing the
all-zero codeword and its neighborhood quadruple system.
In this case we have $S = \Z^{15}_2 \setminus B(Q)$, and
$U = \{ B({\bf x}) : B({\bf x}) \cap B(Q) = \emptyset\}$, where
$B({\bf x}) = \{ {\bf y} : d({\bf x}, {\bf y}) \le 1\}$
and $B(C) = \{B({\bf x}): {\bf x} \in C\}$.
We use the \emph{libexact} software~\cite{libexact} for solving such
instances of the exact cover problems.
In the search we could in fact have made use of the fact that
all 1-perfect binary codes are self-complementary---in other words, the
all-one word is always in the kernel---but this would not have had any
practical significance as the search was rather fast.

Since the all-zero word and its neighborhood quadruple system contain
$141$ of the $2\,048$ codewords, $1907$ new codewords are needed.
Searching for these was a remarkably easy computational task;
on average the search trees in which codes were found had 1978 nodes
and those in which no codes were found had 3 nodes.

\section{Isomorph rejection}\label{sec:isom}

The general framework by McKay~\cite{McKay98} was used to carry out
isomorph rejection, although a less sophisticated method would 
have sufficed in this work.

Recall that we augment a Steiner quadruple system $Q$ to an extended
1-perfect code $C$. We accept $C$ if it passes the following two tests;
otherwise it is rejected. First we require that $C$ shall be the minimum
(with respect to some practically computable total order
of codes) under the action of $\Aut(Q)$. Second, we compute the
canonical equivalence class representative $c_E(C)$, consider
$\pi, {\bf x}$ for which $\pi(C + {\bf x}) = c_E(C)$ and require that
${\bf x}$ and ${\bf 0}$ are on the same $\Aut(C)$ orbit
(we define $c_E$ so that ${\bf x} \in C$ always holds).

When the extended 1-perfect codes have been classified, classifying
the 1-perfect codes is straightforward. All 1-perfect codes are obtained
by puncturing the extended codes, and the resulting 1-perfect codes
are equivalent if and only if they are obtained by puncturing
the same extended code at coordinates which are in the same
orbit of the automorphism group.

A complete classification of the $(14, 1\,024, 3)$ codes is obtained similarly,
since each such code is obtained by shortening a unique (up to equivalence)
1-perfect code of length $15$; this result was proved by
Blackmore~\cite{Blackmore99}. Although a code can be shortened at any
coordinate in two
ways, by selecting the codewords with $0$ or $1$ in a certain coordinate,
both selections lead to equivalent codes. This follows from the fact
that every 1-perfect binary code is self-complementary.

Furthermore we note that any $(15, 1\,024, 4)$ code is obtained by extending
a $(14, 1\,024, 3)$ code with a parity bit. Hence all such codes are
obtained by shortening and extending a perfect code, or equivalently
removing all words of chosen parity. As the perfect codes are
self-complementary, we get (up to equivalence) same code by chosing
either odd or even parity. As this mapping is reversible, we conclude
that there is one-to-one correspondence between equivalence classes of
$(15, 2\,048, 3)$ codes and equivalence classes of $(15, 1\,024, 4)$
codes, and in both cases their number is $5\,983$.

Let $C$ be a $(15, 1\,024, 4)$ code $C$ and let $C'$ be the corresponding
perfect code. The group $\Aut(C')$ contains $\Aut(C)$ as a subgroup,
and $\Aut(C')$ has one more generator than $\Aut(C)$, namely the all-one
codeword. Accordingly $|\Aut(C')| = 2|\Aut(C)|$.

We still have to describe an algorithm for canonical labeling. The most
straightforward approach of using the general purpose isomorphism
\emph{nauty}~\cite{nug} is rather slow on codes as large and regular
as the $(16, 2\,048, 4)$ codes; this was also noted by
Phelps~\cite{Phelps00}. Hence a tailored approach is necessary.
The method presented below has a lot in common with the one
described in~\cite{Phelps00}. An alternative method based on minimum
distance graphs would also work~\cite{MOPS09}, cf.~\cite{PL99}.

A \emph{triangle} consists of $3$ codewords with mutual distance $4$.
Triangles constitute an easily computable and rather sensitive
invariant of Steiner quadruple systems. 
Distinguishing the isomorphism classes of the neighborhood quadruple
systems of a code also constitutes an invariant of the extended 1-perfect codes.
These two invariants turned out to be useful for speeding up our computations.

A canonical isomorphism class representative $c_I(C)$ for a code $C$
can be computed by using \emph{nauty} to label a corresponding graph
canonically. Moreover, \emph{nauty} computes generators of the group $\Sym(C)$.
Canonical equivalence class
representative can be defined as
$c_E(C) = \min \{c_I(C + {\bf x}) : {\bf x} \in C\}$,
where the minimum is again taken with respect to some practically computable
total order of codes.

Note that two codewords, ${\bf x}$ and ${\bf y}$, are in the same orbit
of $\Aut(C)$ if and only if $c_I(C + {\bf x}) = c_I(C + {\bf y})$.
Because of this, if we know that ${\bf x}$
and ${\bf y}$ are in the same orbit, and $c_I(C + {\bf x})$ has
been computed, then there is no need to compute
$c_I(C + {\bf y})$. Also if $c_I(C + {\bf x}) = c_I(C)$, then \emph{nauty}
yields a permutation $\pi$ such that $\pi(C + {\bf x}) = C$. The
pairs $(\pi, {\bf x})$ are coset representatives of $\Aut(C)$ with
respect to $\Sym(C)$, so we get generators of the group $\Aut(C)$.

\section{Results}\label{sec:results}

There are exactly $2\,165$ inequivalent extended 1-perfect codes of
length $16$, $5\,983$ inequivalent 1-perfect codes of length $15$,
$38\,408$ shortened 1-perfect codes of length $14$ and
$5\,983$ $(15, 1\,024, 4)$ codes.
The orders of the automorphism groups of the codes are presented
in Tables~\ref{tbl:extautorder}, \ref{tbl:perfautorder} and
\ref{tbl:shortautorder}. As noted in Section~\ref{sec:isom}, the
order of the automorphism group of a $(15, 1\,024, 4)$ code is half of the
order of the automorphism group of the corresponding perfect code.

The codes have been made available in electronic form by including them in the source of the arXiv version of this paper. Downloading \url{http://arxiv.org/e-print/0806.2513v3} and uncompressing it with {\tt gunzip} and {\tt tar} yields the files {\tt perfect15} and {\tt extended16}.

Only $15\,590$ of the $1\,054\,163$ nonisomorphic $S(3,4,16)$ can
be augmented to a 1-perfect code, and the total number of extensions
is $22\,814$. The computationally intensive part
of this result was the earlier classification of $S(3,4,16)$,
which required several years of CPU time, while all searches
described in this paper took only a couple of hours of CPU time.

A detailed study of the properties of the classified codes will appear
in a second part of this article~\cite{OPP09}.

\begin{table}[h]
\begin{center}
\caption{Automorphism groups of $(16, 2\,048, 4)$ codes}
\label{tbl:extautorder}
\begin{tabular}{rrrrrr}\hline
$|\Aut(C)|$ & \# & $|\Aut(C)|$ & \# & $|\Aut(C)|$ & \# \\ \hline
   128 &  11 &   5\,376 &   1 &      196\,608 & 6 \\
   192 &   5 &   6\,144 &  23 &      262\,144 & 3\\
   256 & 105 &   8\,192 & 174 &      344\,064 & 1\\
   384 &   9 &  10\,752 &   2 &      393\,216 & 3\\
   512 & 377 &  12\,288 &  22 &      524\,288 & 2\\
   672 &   2 &  16\,384 & 103 &      688\,128 & 1\\
   768 &  19 &  24\,576 &  12 &      786\,432 & 2\\
1\,024 & 416 &  32\,768 &  47 &   1\,572\,864 & 3\\
1\,344 &   1 &  43\,008 &   2 &   2\,359\,296 & 1\\
1\,536 &  21 &  49\,152 &  18 &   2\,752\,512 & 1\\
1\,920 &   1 &  61\,440 &   1 &   3\,145\,728 & 1\\
2\,048 & 394 &  65\,536 &  33 &   5\,505\,024 & 2\\
2\,688 &   1 &  86\,016 &   3 &   6\,291\,456 & 1\\
3\,072 &  18 &  98\,304 &  12 & 660\,602\,880 & 1\\
4\,096 & 298 & 131\,072 &   6 \\
\hline
\end{tabular}
\end{center}
\end{table}

\begin{table}[h]
\begin{center}
\caption{Automorphism groups of $(15, 2\,048, 3)$ codes}
\label{tbl:perfautorder}
\begin{tabular}{rrrrrr}\hline
$|\Aut(C)|$ & \# & $|\Aut(C)|$ & \# & $|\Aut(C)|$ & \# \\ \hline
  8 &      3 &     512 & 1\,017 &      24\,576 & 7  \\
 12 &      3 &     672 &      3 &      32\,768 & 8  \\
 16 &      5 &     768 &     32 &      43\,008 & 4  \\
 24 &     10 &  1\,024 &    697 &      49\,152 & 10 \\
 32 &    138 &  1\,536 &     17 &      65\,536 & 5  \\
 42 &      2 &  2\,048 &    406 &      98\,304 & 1  \\
 48 &     12 &  2\,688 &      1 &     131\,072 & 1  \\
 64 &    542 &  3\,072 &     37 &     172\,032 & 1  \\
 96 &     22 &  3\,840 &      1 &     196\,608 & 5  \\
120 &      1 &  4\,096 &    202 &     344\,064 & 2  \\
128 & 1\,230 &  5\,376 &      4 &     393\,216 & 2  \\
192 &     18 &  6\,144 &     35 &     589\,824 & 1  \\
256 & 1\,319 &  8\,192 &     94 & 41\,287\,680 & 1  \\
336 &      3 & 12\,288 &      7 \\
384 &     30 & 16\,384 &     44 \\
\hline
\end{tabular}
\end{center}
\end{table}

\begin{table}[h]
\begin{center}
\caption{Automorphism groups of $(14, 1\,024, 3)$ codes}
\label{tbl:shortautorder}
\begin{tabular}{rrrrrr}\hline
$|\Aut(C)|$ & \# & $|\Aut(C)|$ & \# & $|\Aut(C)|$ & \# \\ \hline
  1 &      5 &    168 &      1 &      8\,192 & 80 \\
  2 &     75 &    192 &     80 &     12\,288 & 18 \\
  3 &      8 &    256 & 4\,392 &     16\,384 & 14 \\
  4 &    425 &    336 &      5 &     21\,504 &  1 \\
  6 &     39 &    384 &    114 &     24\,576 & 15 \\
  8 & 1\,162 &    512 & 2\,469 &     32\,768 & 14 \\
 12 &     56 &    768 &     30 &     49\,152 &  1 \\
 16 & 3\,465 & 1\,024 & 1\,346 &     65\,536 &  1 \\
 21 &      4 & 1\,344 &      1 &     86\,016 &  1 \\
 24 &     39 & 1\,536 &     54 &     98\,304 &  2 \\
 32 & 7\,311 & 2\,048 &    527 &    172\,032 &  1 \\
 48 &     59 & 2\,688 &      6 &    196\,608 &  2 \\
 64 & 9\,068 & 3\,072 &     55 & 1\,376\,256 &  1 \\
 96 &     49 & 4\,096 &    222 & \\
128 & 7\,172 & 6\,144 &     18 & \\
\hline
\end{tabular}
\end{center}
\end{table}

\section{Consistency check}\label{sec:check}

To get confidence in the results, we performed a consistency check
similar to the one used, for example, in~\cite{sqs16}. In this check we count the
total number of codes in two different ways and ensure
that the results agree.

First we consider the set $\mathcal{C}$ of equivalence class representatives
obtained in the classification. By the orbit-stabilizer theorem, the
total number of extended 1-perfect codes is
\[
\sum_{C \in \mathcal{C}} \frac{16! \cdot 2^{16}}{\Aut(C)},
\]
where $16! \cdot 2^{16}$ is the order of the wreath product group acting
on the codes.

Let $\mathcal{Q}$ consist of the representative Steiner quadruple systems,
and let $E(Q)$ be the number of all extended 1-perfect codes obtained by
augmenting $Q$. Applying the orbit-stabilizer theorem, we get the expression
\[
\frac{1}{2\,048} \sum_{Q \in \mathcal{Q}} \frac{16! \cdot 2^{16} \cdot E(Q)}
{\Aut(Q)},
\]
where the division by $2\,048$ is necessary since each code is counted once
for each codeword. Both formulas yield the same result,
$2\,795\,493\,027\,033\,907\,200$. Similarly we also counted the
1-perfect codes and shortened 1-perfect codes in two different ways;
their number is $1\,397\,746\,513\,516\,953\,600$.
Indeed, there are twice as many extended 1-perfect codes as there are
1-perfect codes, since each 1-perfect code admits two extensions: one with
even parity bit and one with odd. Similarly, we get a bijection from
the 1-perfect codes to shortened 1-perfect codes if we shorten each code
by taking, for instance, the codewords with value $0$
in coordinate $15$ and removing that coordinate. Thus there are equally
many 1-perfect codes and shortened 1-perfect codes.

\def\cprime{$'$}

\end{document}